# HiveLink – IoT based Smart Bee Hive Monitoring System

**Ajwin Dsouza**[*], **Aditya P**[**], **Sameer Hegde**[**]

[*] St. Joseph Engineering College, Mangaluru
[**] Canara Engineering College, Benjanapadavu

***Abstract-*** HiveLink—the IoT-based Smart Bee Hive Monitoring System addresses the challenges faced by beekeepers in managing the influence of environmental impact, diseases, and collapse in honey bee colonies. Integrated with advanced sensors, the system monitors temperature, humidity, hive weight, and diurnal cycle. Leveraging IoT technology, the system provides real-time data, remote connectivity, and actionable insights for beekeepers. Monitoring the hive with the system enables early disease detection, proactive interventions, and optimized hive management. Minimizing manual inspections, enhancing productivity, and promoting sustainable practices to mitigate environmental impact and support honey bee populations. Therefore, this system is a demonstration of technology-driven solution to ensure the well-being of bee hives by facilitating data-driven decision-making and contributes to the resilience of beekeeping in the face of diverse challenges.

***Index Terms****- Apis cerana*, IoT, Smart bee hive

## I. INTRODUCTION

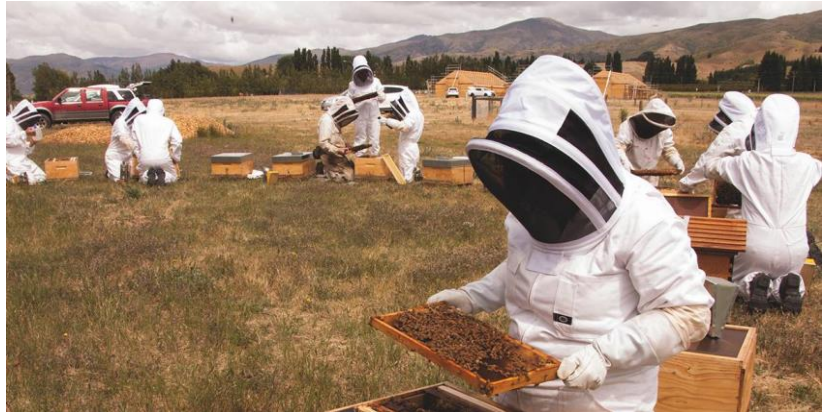

*Figure 1. A photograph showing beekeepers engaged in hive maintenance in a field setting.*
*(Source: https://opmarketing-prod.imgix.net)*

Honey bees play a significant role in agriculture and the world's economy as they are responsible for the pollination of more than one-third of the global food. They also play a crucial role in maintaining healthy ecosystems. However, beekeepers face various challenges in ensuring the well-being and productivity of their honey bee colonies. Environmental factors, including temperature fluctuations and humidity levels greatly influence the behavior and health of honey bees. Diseases and pests pose significant threats to colony health and can lead to colony collapse. Besides these challenges, beekeepers also need to address issues such as swarming—the act of relocation after leaving the existing hive and absconding causing disruptions to hive populations.

To effectively manage these challenges, beekeepers require an efficient system that can provide real-time data on crucial parameters allowing beekeepers to monitor hive conditions and make informed decisions. By promptly detecting changes in the environment conditions or signs of disease outbreaks, beekeepers can take the necessary actions to ensure hive health and productivity.

Such an effective hive monitoring system can also help beekeepers save time and effort by reducing the need for frequent manual inspections. By leveraging advanced technologies, such as the Internet of Things (IoT) and sensor integration, beekeepers can monitor multiple hives simultaneously, making it easier to identify potential issues and implement timely interventions.

Therefore, the importance of beekeeping in supporting pollination and food production necessitates the need for an efficient hive monitoring system. By addressing challenges related to environmental impact, diseases, and other factors such as, swarming and absconding, a well-designed monitoring system can contribute to sustainable beekeeping practices, the preservation of honey bee populations, and the overall health of our ecosystems.

## II. RESEARCH PURPOSE

**2.1. Inefficient Hive Management**
Traditional manual inspections require significant time and effort, limiting beekeeper's ability to monitor hive conditions effectively and respond promptly to issues in large-scale beekeeping operations.

**2.2. Limited Real-Time Insights**
Beekeepers lack real-time access to crucial hive parameters, such as temperature, humidity, hive weight, hindering their ability to make informed decisions and optimize hive conditions.

**2.3. Delayed Detection of Environmental Changes**
Environmental factors, including temperature fluctuations and humidity levels, impact honey bee health and behavior. Without timely detection, beekeepers are unable to implement necessary adjustments to maintain optimal hive conditions.

**2.4. Inability to Monitor Multiple Hives Simultaneously**
Beekeepers with numerous hives face challenges in simultaneously monitoring each hive's status, making it difficult to identify hive-specific requirements and address issues promptly.

**2.5. Lack of Early Disease/Pest Detection**
Viral infections and pests such as Wax Moth, Varroa mites can devastate honey bee colonies if not detected and treated early. Current practices may result in delayed disease/pest identification and limited opportunities for timely intervention leading to colony loss.

**2.6. Swarming and Absconding Events**
Honey bees may exhibit swarming behavior or absconding from their hives, leading to population loss and disruption in hive productivity. Beekeepers require tools to detect and manage these events effectively.

**2.7. Inefficient Resource Allocation**
Without real-time data on hive conditions and honey production, beekeepers may struggle to allocate resources effectively, resulting in suboptimal hive management and reduced productivity.

**2.8. Hive Entrance Vulnerability at Night**
Bees are attracted to light sources at night, which can lead to bees leaving the hive and increase potential loss. Beekeepers, thus, face the challenge of protecting their hives by closing the hive entrance at night.

**2.9. Difficulty in Feeding Supplements**
Providing supplementary nutrition to honey bee colonies can be challenging, especially when determining the appropriate timing and quantity. Beekeepers require assistance in monitoring hive conditions and determining the optimal feeding schedule to support colony health.

**2.10. Hive Theft**
Beehives can be targets for theft. The loss of beehives can result in financial loss and disruption to honey production.

**2.11. Fall-over of Hives**
Hives may be vulnerable to falling or tipping due to tree branches, heavy winds, or other external factors. Such incidents can cause damage to the hive structure, devastate the colony, and impact overall hive stability.

## III. RESEARCH OBJECTIVE

**3.1. Reduce Inspection Time**
Implementation of IoT based Smart Bee Hive Monitoring System increases capabilities to minimize the need for frequent manual inspections of hives. This objective aims to streamline hive management processes, save time for beekeepers, and reduce labor-intensive tasks associated with inspections.

**3.2. Improve Responsiveness to Hive Conditions**
Enable beekeepers to promptly respond to changes in hive conditions by providing real-time data and alerts. This objective aims to empower beekeepers to make informed decisions and implement timely interventions to ensure optimal hive health and productivity.

**3.3. Enhance Data-Driven Decision Making**
Develop analytical tools and visualization techniques to interpret the collected hive data effectively. This objective aims to provide beekeepers with actionable insights, enabling them to make data-driven decisions for hive management, including honey flow prediction and timely feeding of supplements and adjustments based on hive conditions.

## 3.4. Autonomous Hive Entrance Closing at Night
Integrate the monitoring system to autonomously close hive entrances during night and open during day-light period. This objective aims to prevent bees from exiting the hive when attracted to external light sources during darkness, reducing the risk of bees leaving the hive and ensuring their safety and well-being.

## 3.5. Theft Detection
Incorporate theft detection mechanisms into the hive monitoring system. This objective aims to protect beehives from unauthorized access and theft incidents, providing beekeepers with peace of mind and ensuring the security of their valuable assets.

## 3.6. Feeding Indication
Implement notifications to assist beekeepers in determining the appropriate timing and quantity for supplement feeding. This objective aims to address the difficulty in feeding supplements by providing timely guidance and reminders based on hive conditions and nutritional needs.

## 3.7. Early Detection of Disease Outbreaks
Implement disease monitoring algorithms and anomaly detection techniques to enable early identification of potential disease outbreaks. This objective aims to facilitate targeted treatments, to prevent the spread of diseases and mitigate their impact on hive health.

## 3.8. Evaluate System Performance and User Satisfaction
Conduct rigorous testing and gather feedback from beekeepers to assess the performance, reliability, usability, and user satisfaction of the IoT-based hive monitoring system.

## 3.9. Implementation of API for data accessibility
The data sourced from Beehives holds significant value, as it can serve as a valuable resource for surveys and research conducted by governmental bodies and agricultural universities. This data not only provides insights crucial for effective hive management but also offers a robust foundation for comprehensive studies aimed at advancing beekeeping practices and supporting broader agricultural research initiatives.

## IV. ARCHITECTURE OF HIVELINK

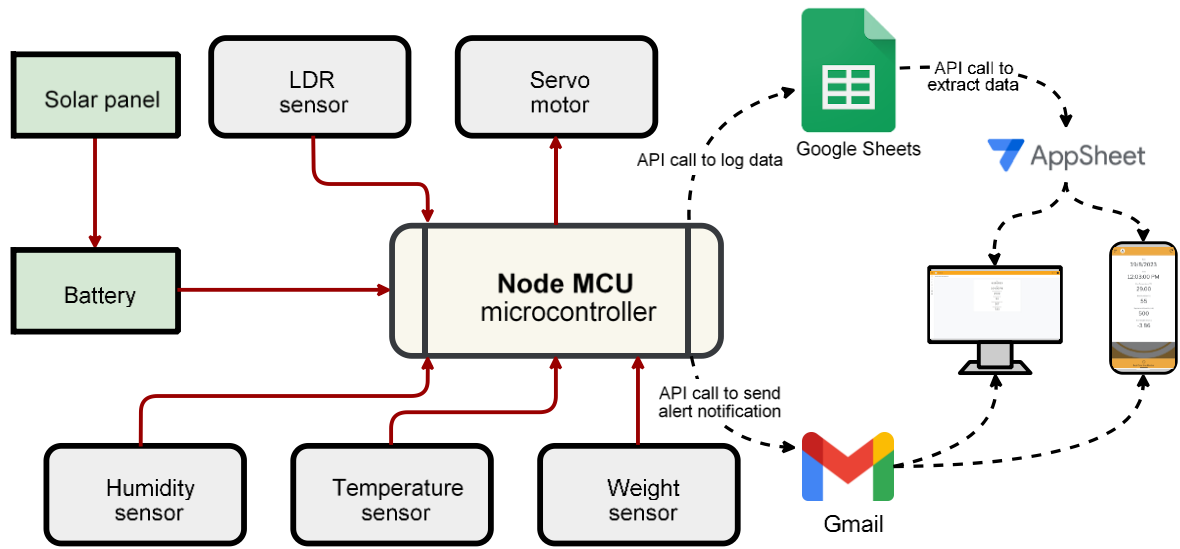

*Figure 2. Block diagram of HiveLink module*

The overall architecture of the system consists of several components working together to collect, transmit, and analyze hive data. The following are the key components involved in the architecture:

## 4.1. Hive Sensors
The system incorporates various sensors to monitor important parameters. These sensors include temperature and humidity sensors (DHT sensor), weight sensors (10kg load cells) are integrated with Load Cell Amplifier, liquid level sensor and light sensor (LDR sensor). The sensors are strategically placed within the hive to collect accurate and reliable data.

## 4.2. Servo Motor
The system incorporates a servo motor as an output device for controlling the hive entrance gate. The microcontroller can receive commands from the user interface to open or close the gate.

### 4.3. Microcontroller
Node MCU acts as the central processing unit of the system. It interfaces with the sensors to collect data and performs necessary computations and operates Servo. The microcontroller also facilitates communication with the Google Spreadsheet for data transmission.

### 4.4. Connectivity
The system utilizes Internet of Things (IoT) connectivity to establish a connection between the hive monitoring system and Google Spreadsheet through Google Apps Script based API. This is achieved through Wi-Fi technology. The connectivity allows seamless data transmission and remote access to hive data.

### 4.5. Cloud Platform
HiveLink utilizes Google Apps Script which is a cloud based scripting platform to store and process the collected data. The cloud platform provides a secure and scalable environment for data management. It enables real-time data analysis, storage, and visualization for beekeepers.

### 4.6. Beekeeper Interface
The system includes Google AppSheet application software, a user-friendly interface that allows beekeepers to monitor and manage hive data remotely. This interface can be accessed through web or mobile application. It provides real-time data visualization, alerts, and historical data analysis, enabling beekeepers to track hive conditions and respond promptly to any issues.

### 4.7. Alert Notification Interface
Alert notifications are facilitated via the IFTTT platform through API, ensuring seamless communication through email notifications. This integrated system enables the timely dissemination of alerts, allowing beekeepers to stay informed about critical hive conditions and take prompt actions as necessary.

## V. SYSTEM DESIGN AND IMPLEMENTATION

### 5.1. Hardware Setup

#### 5.1.1. Circuit implementation

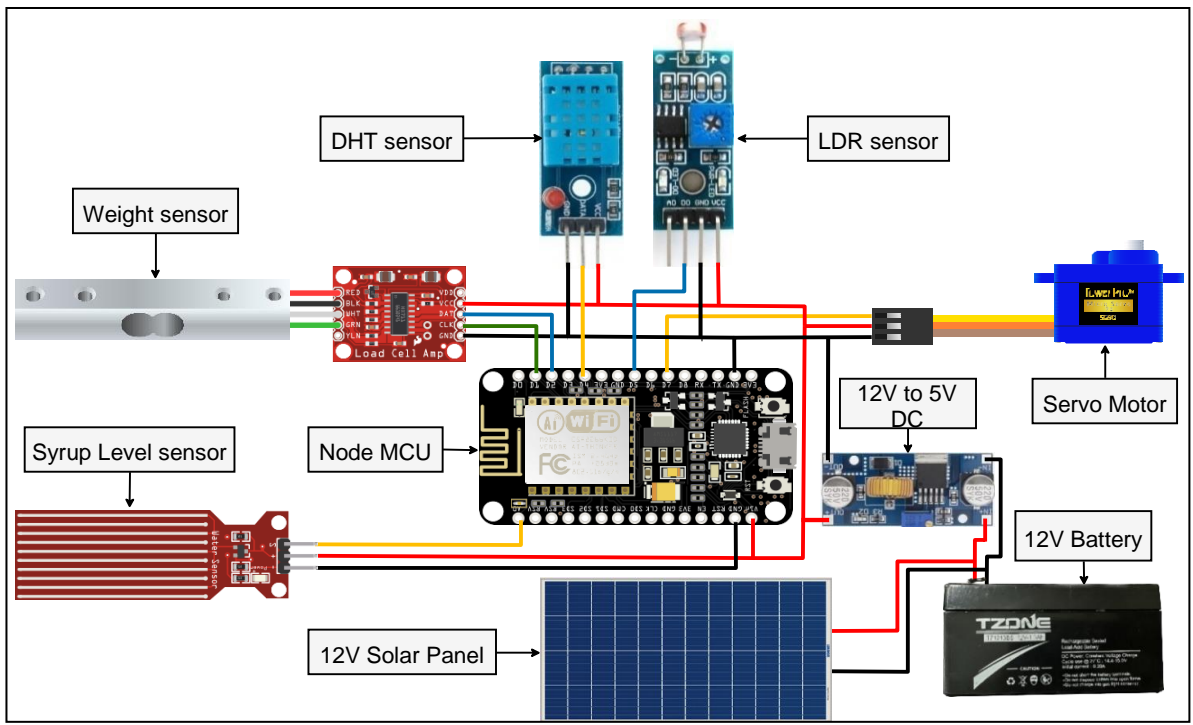

*Figure 3. Circuit Implementation of HiveLink Module*

In HiveLink module, the hardware components are interconnected to enable efficient hive management. The system is powered by a 12V solar panel placed on top of the Hive, which charges a 12V sealed lead-acid battery. The battery is connected to an XL4015 DC-DC buck converter, which regulates the voltage to 5V. This 5V output is then supplied to the NodeMCU microcontroller, as well as the sensors and servo motor.

The microcontroller used in the system is NodeMCU, which acts as the central processing unit. The sugar supplement level sensor is connected to pin A0 of the microcontroller, allowing for accurate monitoring of the sugar supplement level in the hive.

The load cell, along with the HX711 amplifier, clock and data pins is connected to the D1 and D2 pins of the microcontroller respectively, enabling precise measurement of hive weight.

For environmental monitoring, the DHT sensor's data pin is connected to pin D4 of the microcontroller, providing real-time temperature and humidity readings. The LDR sensor module's D0 pin is connected to pin D5 of the microcontroller, allowing for monitoring of light intensity within the hive.

To control the hive entrance gate, a servo motor is employed. The signal pin of the servo motor is connected to pin D7 of the microcontroller. This enables the microcontroller to control the rotation of the servo motor, autonomously opening or closing the hive entrance based on predetermined conditions.

The power supply from the battery and the connections between the microcontroller, sensors, and servo motor are appropriately managed to ensure stable operation and reliable data acquisition. The 5V output from the DC-DC buck converter powers the NodeMCU microcontroller and the various sensors, allowing for seamless data collection and transmission.

In summary, the Smart Bee Hive Monitoring System utilizes a 12V solar panel and a sealed lead-acid battery for power supply. The XL4015 DC-DC buck converter regulates the voltage to 5V, which is distributed to the NodeMCU microcontroller, sugar supplement level sensor, load cell with HX711 amplifier (with clock connected to D1 and data connected to D2), DHT sensor, LDR sensor module, and servo motor. These hardware components and connections enable accurate monitoring of hive conditions, weight measurement, environmental sensing, and autonomous control of the hive entrance, facilitating efficient hive management and enhancing beekeeper's decision-making processes.

*5.1.2. Module assembly*

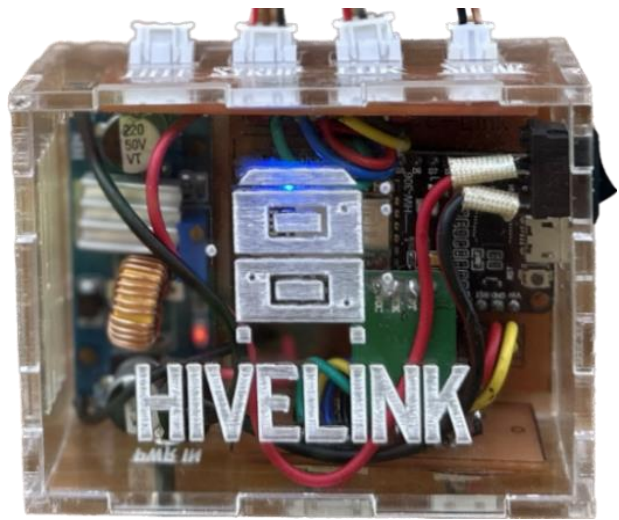

*Figure 4. photograph of HiveLink Module Assembly*

The module assembly for the Smart Bee Hive Monitoring System involves integrating the NodeMCU microcontroller, buck converter (XL4015), HX711 amplifier, and other components onto a PCB (Printed Circuit Board). The assembly ensures proper connections and organized placement of all the components. The NodeMCU microcontroller, buck converter, and HX711 amplifier are securely mounted on the PCB. The sensors, including the DHT sensor, LDR sensor module, and syrup level sensor, are connected to the respective pinouts on the PCB. The servo motor is also connected to a designated pinout for actuator control. The power supply connections, including the 12V solar panel, sealed lead acid battery, and buck converter, are established on the PCB. The entire module assembly is housed within a transparent acrylic sheet enclosure, providing physical protection and a tidy arrangement of wires and components.

*5.1.3. Placement of HiveLink Setup on the Beehive*

The placement of sensors, servo, and module on the hive is crucial for effective monitoring and control. Here is a description of their placement:

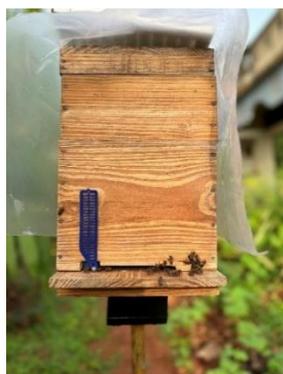 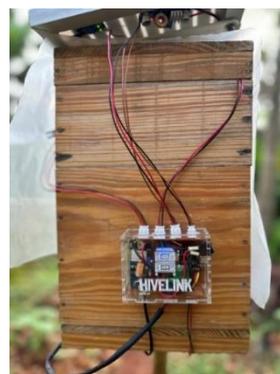

*Figure 5. Front view of HiveLink enabled Beehive*   *Figure 6. Back view of HiveLink enabled Beehive*

The servo motor is mounted near the hive entrance for the gate mechanism (Figure 5). It is securely attached to the hive structure, allowing smooth opening and closing of the hive entrance. The servo motor's movement is aligned with the hive entrance to control access and prevent exit of bees at night.

The HiveLink module, which includes the microcontroller and associated circuitry, is typically placed behind the hive (Figure 6), in a weatherproof enclosure. It serves as the central unit for data collection, processing, and communication with the monitoring system.

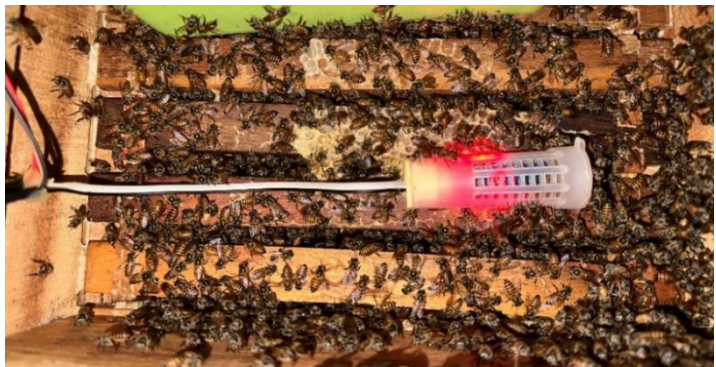

*Figure 7. Photograph depicting Placement of DHT sensor*

The DHT sensor is enclosed inside a porous container which is positioned inside the hive, near the brood area or in a central location, ensuring accurate measurement of temperature and humidity levels within the hive.

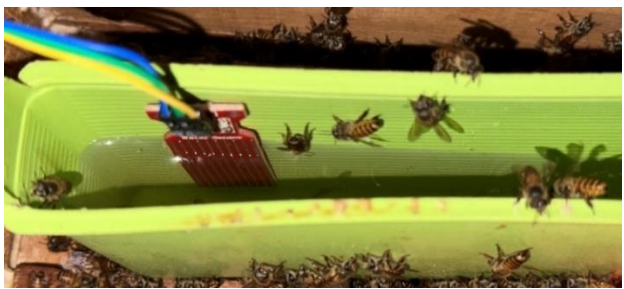

*Figure 8. Photograph of Syrup Level Sensor placed inside supplement Feeder*

The sugar syrup level sensor is installed within the supplement feeder inside the Beehive (Fig 4.1.6). It is positioned to detect the level of the sugar supplement accurately in mL. The sensor's probe or electrode is carefully inserted into the syrup container, ensuring proper contact for reliable measurements.

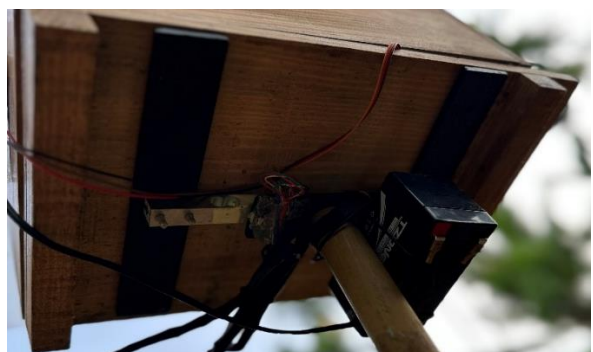

*Figure 9. Load cells connected to form H shaped support for the hive.*

Two 10 kg load cells are placed underneath the hive, providing support in an H-shaped structure. It is positioned to evenly distribute the weight of the hive and accurately measure any changes in hive weight. The load cell is securely fixed to the hive bottom, ensuring stability and precise weight monitoring, Alongside Battery pack is placed below the H structure and its weight is not measured by Load cells.

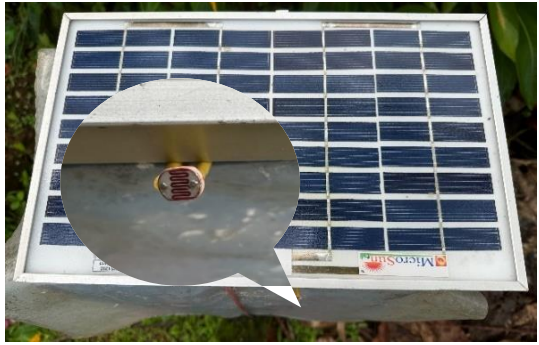

*Figure 10. Photograph of top view of Hive depicting LDR sensor on the Solar Panel*

The LDR sensor module is placed outside the hive, mounted to the Solar Panel, which is placed above the Hive. It is positioned to measure the ambient light levels surrounding the hive.

**5.2. Software Integration**

The software integration strategy has been adapted to enhance flexibility and functionality, utilizing the NodeMCU platform and various interconnected tools. This multifaceted approach encompasses distinct phases and tools, each contributing to a comprehensive hive monitoring and management solution. The integration process included the following steps:

*5.2.1. Programming using Arduino IDE*

*Figure 11. Arduino IDE platform*

In this phase, the NodeMCU is programmed using the Arduino Integrated Development Environment (IDE). Basic sensor integration, Google Apps Script linking and IFTTT notification triggering code segments are developed to ensure fundamental hive monitoring system operations. This programming facilitates data collection from various sensors and lays the foundation for subsequent stages.

*5.2.2. Spreadsheet Syncing and Data Transfer via Google Apps Script*

*Figure 12. Google Apps Script Development Platform*

To ensure seamless data transmission and storage, the system integrates with Google Sheets. Custom code and Google Apps Script facilitate the efficient transfer of data collected by the NodeMCU to a designated Google Sheets document. This integration forms a vital bridge between data acquisition and analysis, leveraging REST APIs to streamline the process.

*4.2.3. User Interface Development with Google App Sheets*

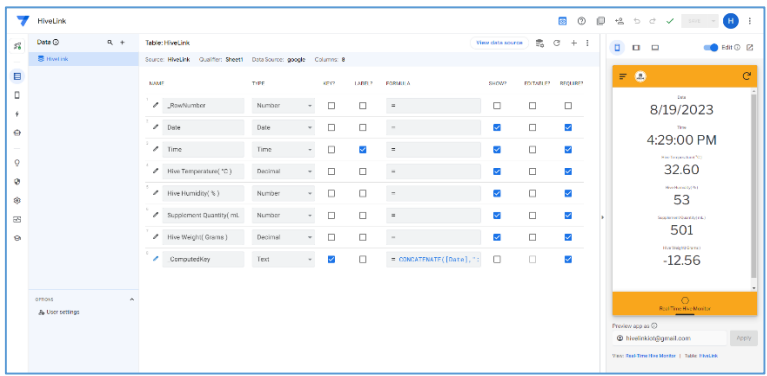

*Figure 13. Google Apps Script Development Platform*

Google App Sheets is employed to create an intuitive user interface, empowering beekeepers with real-time insights. Through interactive graphs and real-time Beehive Monitoring, Beekeepers can conveniently monitor hive conditions. This user-friendly interface facilitates effective decision-making by presenting data in an easily understandable format.

*4.2.4. IFTTT Integration for Alert Notifications*

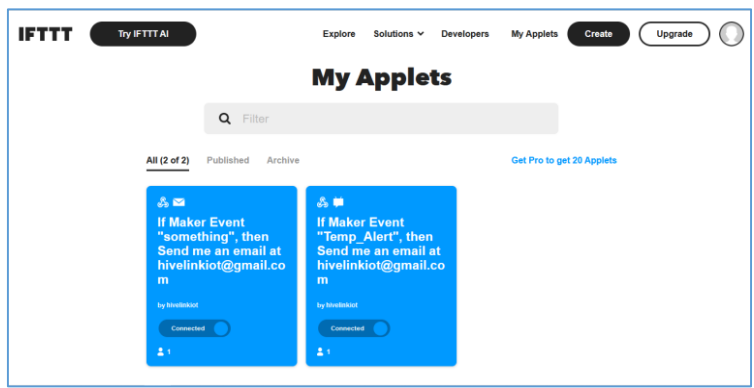

*Figure 14. IFTTT Development Platform*

The IFTTT platform is harnessed to enable automated alert notifications. Predefined triggers, based on specific parameters, prompt IFTTT to promptly send email notifications to beekeepers. This proactive alert system ensures swift awareness of critical hive conditions, leveraging APIs for seamless communication.

## VI. RESULTS AND DISCUSSION

The implementation of the Smart Bee Hive Monitoring System yields significant outcomes, enhancing hive management practices and fostering collaboration among various stakeholders.

**6.1. Beekeeper's Interface**

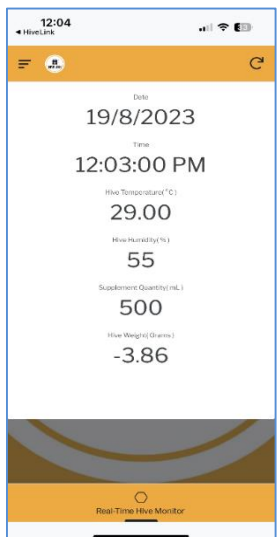
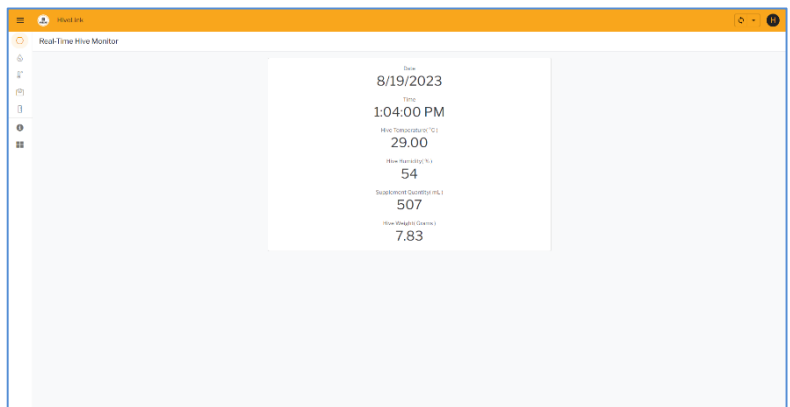

*Figure 15. HiveLink Mobile application*     *Figure 16. HiveLink Website*

The user-friendly interface, developed using Google App Sheets, empowers beekeepers with real-time insights into hive conditions. The interface showcases data through interactive graphs and real-time value displays. This accessible and comprehensible presentation of data aids beekeepers in making informed decisions and optimizing hive management strategies.

**6.2. Google Spreadsheet Data Utilization:**

*Figure 17. Google Spreadsheet having HiveLink data logged*

Data collected from the hives is systematically logged and organized within Google Sheets through custom code and Google Apps Script. This data repository serves as a central hub for hive information, ensuring easy access and efficient management.

**6.3. Contribution to Government Bodies, Agricultural Colleges, and Researchers**

The amassed data holds immense potential for survey and research activities. Government bodies can leverage this data through APIs to gain insights into bee colony health, behavior patterns, and environmental conditions. Agricultural colleges can integrate this real-world data into their curriculum to educate future beekeepers and researchers. Researchers gain access to a wealth of information for conducting studies on hive dynamics, disease prevalence, and sustainable beekeeping practices.

The Smart Bee Hive Monitoring System not only streamlines hive management but also fosters a collaborative ecosystem wherein beekeepers, educational institutions, government bodies, and researchers collectively contribute to the advancement of beekeeping practices and environmental conservation. This integrated approach ensures the sustainability and well-being of honey bee colonies while facilitating scientific exploration and knowledge dissemination.

**6.4. Review on analysed data**

The analysis of data collected from the sensors provides valuable insights into various aspects of hive management. Following is a summary of the key observations and recommendations based on the collected data:

*6.4.1. Disease Prediction*

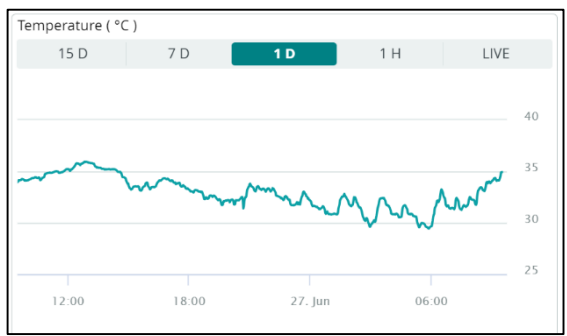
*Figure 18. Temperature graph of Hive under normal conditions*

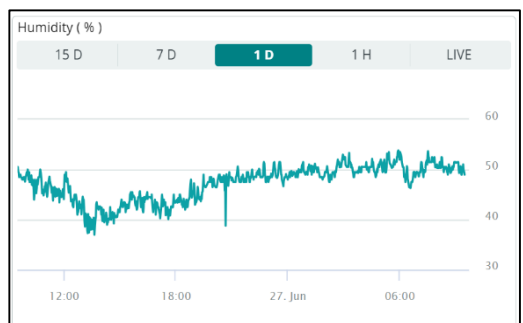
*Figure 19. Humidity graph of Hive under normal conditions*

The temperature and humidity data help in monitoring the hive's health status. Normal conditions are indicated by the bees maintaining a temperature of 30-32 °C and humidity levels between 50–60%. Deviations from these normal conditions can be indicative of potential disease outbreaks. By monitoring temperature and humidity patterns, beekeepers can detect early signs of diseases and take appropriate measures for treatment and prevention [2].

*6.4.2. Absconding*

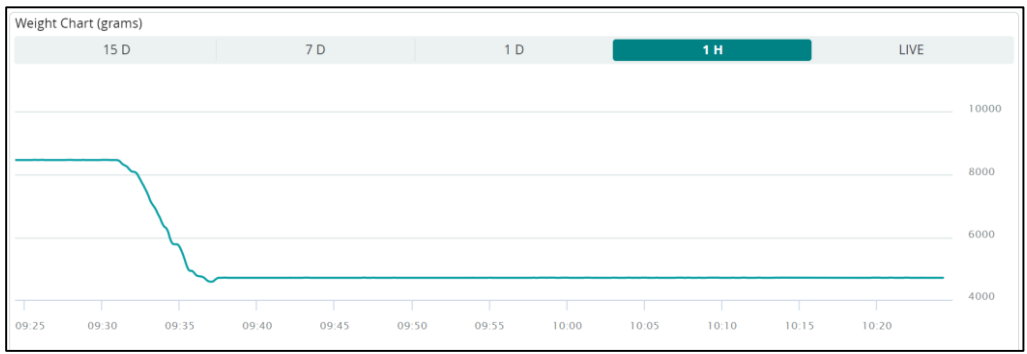

*Figure 20. Weight graph of Hive Depicting Absconding character of Bees.*

Absconding, which refers to the sudden departure of bees from the hive, is identified through significant weight drops of around 1–2 kg. Simultaneously, the temperature and humidity inside the hive return to the ambient levels relatively quickly. Detecting such weight drops and changes in temperature and humidity patterns enables beekeepers to address the underlying issues and respond to absconding.

*6.4.3. Swarming Prediction*

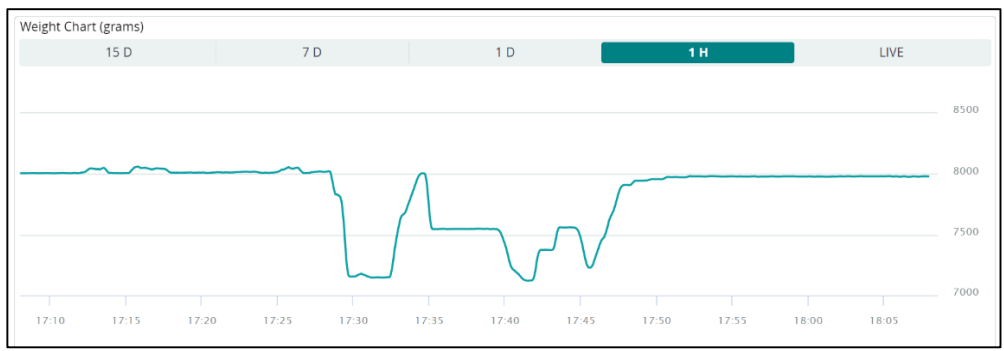

*Figure 21. Weight graph of Hive Depicting Swarming character of Bees.*

Swarming behavior can be predicted based on hive weight observations. As swarming approaches, the hive weight tends to increase. Additionally, the bees exhibit specific behavioral patterns, such as increased flight activity in the late afternoon or evening several days before swarming. Monitoring these weight and behavioral changes allows beekeepers to prepare for swarming events, such as providing extra space or implementing swarm prevention techniques [1].

*6.4.4. Theft / Unauthorized Hive Removal Detection*

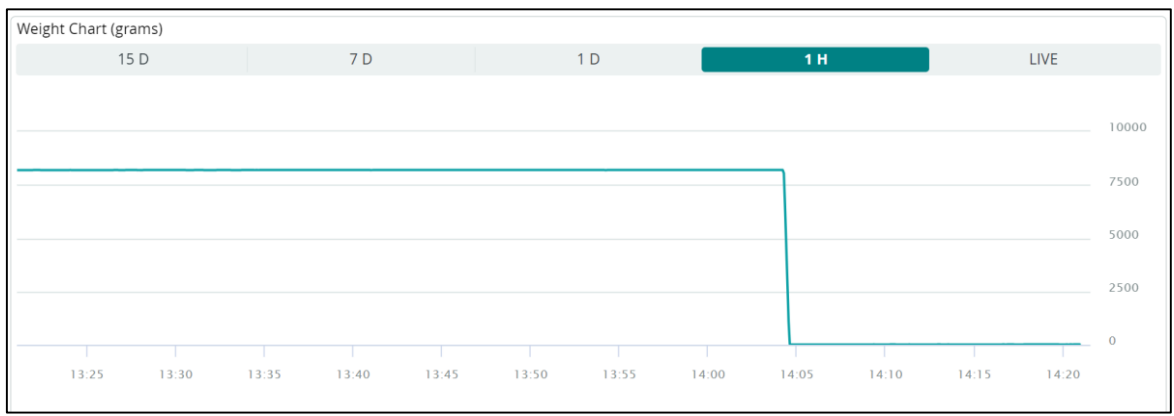

*Figure 22. Weight graph of Hive Depicting Theft/ Unauthorized removal of Hive.*

Theft incidents are detected through a sudden and significant drop in hive weight to zero, indicating the removal of the hive box by unauthorized individuals. This information helps beekeepers take immediate action to address the theft and ensure the safety of their hives.

### 6.4.5. Hive Falls

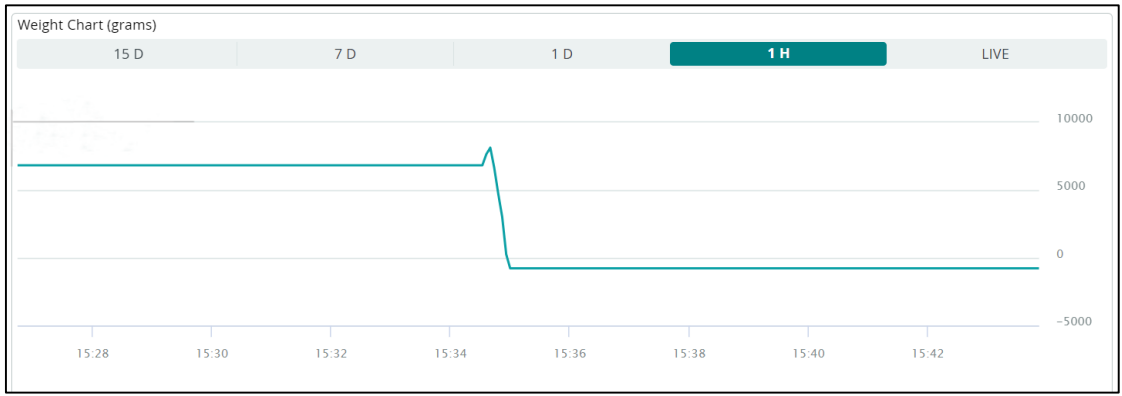

*Figure 23. Weight graph of Hive Depicting fallen state of Hive*

If a hive falls, the weight registered by the load cell becomes negative due to the reverse force of the ground. Monitoring negative weight values alerts beekeepers to the potential hive fall incidents. Prompt intervention can be taken to assess and rectify any damage caused by the fall.

### 6.4.6. Honey Monitoring

Hive weight analysis provides insights into honey production. A constant hive weight combined with a stable bee population weight indicates the absence of significant honey flow. However, the weight increase in a hive can be used to predict honey filling. Typically, under ideal honey flow conditions, it takes 14-15 days for a honey super chamber to fill, with an average increase of 4-5 kg, which is equivalent to 0.2-0.3 Kg per day. This information allows beekeepers to determine the optimal time for honey extraction.

### 6.4.7. Supplement Refilling

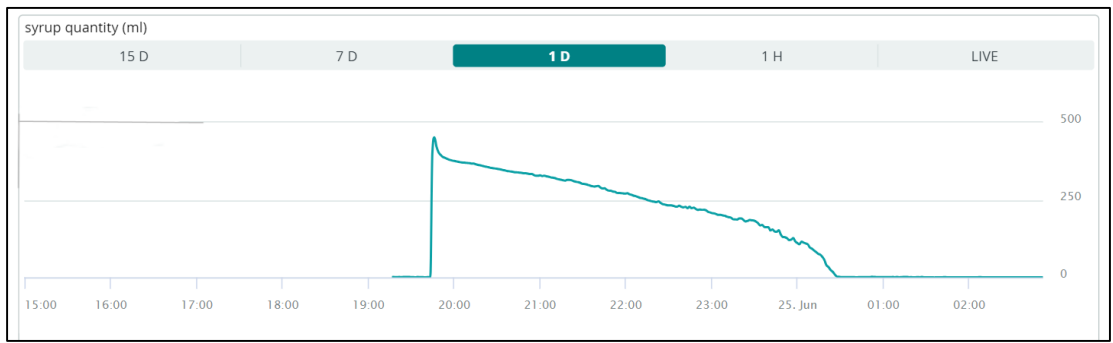

*Figure 24. graph of Supplement consumption in a Hive*

Monitoring the decline in the syrup level helps in planning the next feeding session. By observing the time needed for the bees to consume the supplement and reach zero levels, beekeepers can predict the appropriate time for refilling. This proactive approach ensures that the bees have a continuous supply of the necessary nutrients.

### 6.4.8. Autonomous Gate Control

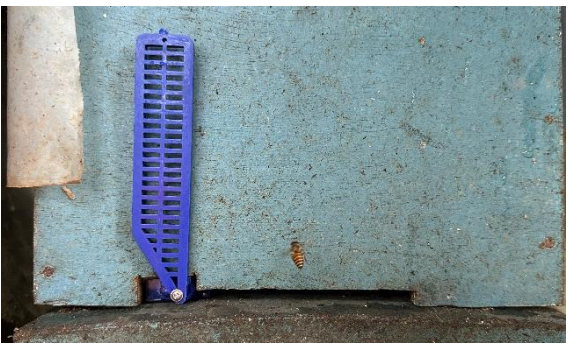 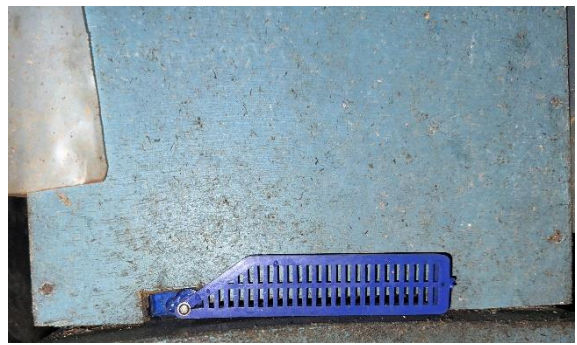

*Figure 25. Hive entrance open during the Day*     *Figure 26. Hive entrance close at night*

The use of an LDR (Light Dependent Resistor) allows for automated gate control. The gate is closed after 7:00 pm, ensuring the bees are protected from external light sources and disturbances. It is then opened at 6 am, allowing the bees to resume scouting and foraging activities.

By analyzing the collected data, beekeepers gain valuable insights into hive conditions, bee behavior, and potential issues. These observations enable them to make informed decisions, implement appropriate interventions, and optimize hive management practices for the well-being and productivity of the honey bee colonies.

## VII. CONCLUSION

In conclusion, HiveLink offers a comprehensive solution for beekeepers to effectively manage their hives and address various challenges. By leveraging IoT technology and sensor integration, the system provides real-time monitoring of temperature, humidity, hive weight, syrup level, and day/night conditions. Through the analysis of collected data, beekeepers can make informed decisions and take proactive measures to ensure the well-being and productivity of honey bee colonies.

The system enables early detection of potential diseases through temperature and humidity monitoring, allowing for timely intervention and prevention. It also facilitates the prediction of swarming events based on hive weight and behavioral patterns, enabling beekeepers to prepare and implement necessary measures. Theft detection capabilities ensure hive security, while the monitoring of hive weight helps identify instances of hive falls and prevent potential damage.

The system's ability to monitor honey filling in the super chamber allows beekeepers to determine the optimal time for honey extraction, optimizing honey production and harvest efficiency. Additionally, the monitoring of syrup levels aids in timely feeding supplement refilling, ensuring the bees receive necessary nutrients.

The integration of an LDR module for autonomous hive entrance control enhances hive protection by automatically closing the entrance at night and opening it in the morning, providing a conducive environment for the bees.

Overall, HiveLink significantly reduces the inspection time required by beekeepers and enhances their ability to respond quickly and effectively to specific hive conditions. This technology-driven solution improves hive management, increases productivity, and promotes the well-being of honey bee colonies. By combining hardware components, data analysis, and informed decision-making, beekeepers can optimize their operations and contribute to the sustainable growth of the beekeeping industry.